\documentclass[aps,prd,preprint,nofootinbib]{revtex4}
\usepackage{amsfonts}
\usepackage{epsfig}
\usepackage{graphicx}
\usepackage{hyperref}
\begin{document}

\title{The Fragmentation Function of Gluon Splitting into P-wave Spin-singlet Heavy
Quarkonium \vspace{0.5cm}}

\author{\bf Gang Hao$^{1}$, YaBing Zuo$^{1,2}$,
Cong-Feng Qiao\footnote{qiaocf@gucas.ac.cn, corresponding
author}$^{1,3}$ \vspace{0.3cm}}

\affiliation{$^1$Department of Physics, Graduate University, the
Chinese Academy of Sciences \\ YuQuan Road 19A, Beijing 100049,
China\\  $^2$Department of Physics, Liaoning Normal University,
Dalian 116029, China\\ $^3$Theoretical Physics Center for Science
Facilities (TPCSF), CAS,\\ YuQuan Road 19B, Beijing 100049, China
\vspace{1.8cm}}

\begin{abstract}\vspace{3mm}
The gluon fragmentation function of $g \rightarrow h_c$ at leading
order of strong coupling constant $\alpha_s$ and typical velocity
$v$  is calculated in the framework of NRQCD, of which the
contributions from both color-singlet and -octet processes are taken
into account and hence the result is infrared safe. It is found that
the fragmentation probability of a high energy virtual gluon to
split into $h_c$ is about $7.1\times 10^{-7}$. The analytic
fragmentation function presented in this work can be employed to the
study of $h_c$ production in large transverse momentum at hadron
colliders, such as Large Hadron Collider(LHC), and is also
applicable for the study of $h_b$ hadronproduction.

\vspace {7mm} \noindent {\bf PACS number(s):} 13.85.Ni, 12.38.Bx,
14.40.Pq
\end{abstract}

\maketitle
%%%%%%%%%%%%%%%%%%%%%%%%%%%%%%%%%%%%%%%%%%%%%%%%%%%%%%%%%%%%%%%%%%%%%%%%%

It is well-known that the fragmentation mechanism dominates the
heavy quarkonium hadroproduction at large transverse momentum
\cite{BraatenLett,Braatenprd50}. It is hence important to obtain the
corresponding fragmentation function in order to properly estimate
the production rate of a specific charmonium state. Fortunately it
was found that these fragmentation functions for heavy quarkonium
production are analytically calculable by virtue of perturbative
quantum chromodynamics(QCD), with limited
universal(phenomenological) parameters.

In past years, Braaten {\it et al.} calculated the fragmentation
functions of gluon to the S-wave charmonium states $\eta_c$ and
$J/\psi$ \cite{BraatenLett,Braatenprd52}. Later, the gluon
fragmentation to polarized $J/\psi$ was obtained by fitting
\cite{qqw}, and the fragmentation function of gluon to $J/\psi$ via
color-octet mechanism was calculated to next-to-leading(NLO) order
accuracy \cite{bl}. The fragmentation functions of gluon to P-wave
spin-triplet charmonium states $\chi_{cJ}$ were found infrared cut
dependent \cite{ma1}, and after including the color-octet mechanism
they then become infrared safe \cite{Braatenprd50}. The polarized
fragmentation functions of gluon to polarized $\chi_{cJ}$ were also
obtained \cite{ChoWise2}. The fragmentation functions for gluon to
D-wave spin-singlet and -triplet were presented in
\cite{ChoWise1,QiaoYuanZhao}. The fragmentation functions of heavy
quark splitting into S-wave charmonium states were calculated in
\cite{frag5,frag6,frag7}, and the corresponding calculations for
higher excited states, the P-wave and D-wave states, were given in
Refs.\cite{ma2,frag8,frag9,cy}. Notice that since in this letter we
are not going to give a complete review of the calculation of
quarkonium fragmentation functions, readers who are interested in
the details on this respect should refer to recent review articles,
like \cite{bes3}.

From above brief introduction, one may notice that the fragmentation
function of gluon splitting into the spin-singlet P-wave state $h_c$
is still absent. Because the production rate for $h_c$ is low and
the decay modes of $h_c$ are obscure, which hinder experimenters to
measure it, for a long time the study of $h_c$ hadroproduction in
large transverse momentum is not so urgent and hence the
fragmentation function for it. However, in recently the $h_c$ is
observed in experiment \cite{e835,cleo1,cleo2,cleo3}, and next
people may further study its detailed natures. With the development
of relevant experiments, people expect that more data on $h_c$ will
be collected, like in hadron collision, the LHC for example, in the
near future \cite{qy,Sridhar,O8value2}. In this sense, the
fragmentation function of gluon splitting into $h_c$ is necessary
and valuable. In this paper we calculate it at the leading order of
$\alpha_s$ and $v$ in the framework of non-relativistic QCD(NRQCD)
\cite{nrqcd1,nrqcd2}, including contributions from both
color-singlet and color-octet.

Generally speaking, heavy quarkonium hadroproduction via
fragmentation mechanism involves both short-distance and
long-distance effects. The differential cross section can be
expressed as a convoluted form \cite{APequation},
\begin{eqnarray}
d \sigma_{H} ( p ) \approx \sum_i\int_0^1 d z\, d\widehat{\sigma}_i
( p /z,\mu ) D_{i \rightarrow H } ( z,\mu )\; .
\end{eqnarray}
This is a factorized formula, in which the short- and long-distance
effects have been separated. $d\widehat{\sigma}$ is the cross
section for parton $i$ production, and is perturbative QCD(pQCD)
calculable due to the energetic parton and large momentum transfer.
$D_{i\rightarrow H}(z, \mu)$ is a process-independent fragmentation
function, which describes how parton $i$ with invariant mass $\mu$
splitting into hadron $H$ with light-cone momentum fraction $z$. For
light hadrons, like $\phi$ and $\pi$, the fragmentation processes
involve low energy scales, such as $\Lambda_{QCD}$ and
$m_{\phi,\pi}$, the fragmentation function is therefore a
non-perturbative quantity and cannot be computed within the scope of
pQCD. However, for the fragmentation process of a parton splitting
into heavy quarkonium, though there will be some non-perturbative
effects due to hadronization, the production of a heavy quark is a
short-distance process and hence is pQCD accessible. It turns out
therefore that the fragmentation function for heavy quarkonium
production can be further factorized \cite{Braatenprd52}.

In the framework of non-relativistic QCD, the fragmentation function
of a virtual gluon splitting into heavy quarkonium $H$ reads
\begin{eqnarray}\label{factor-formula}
D_{g\rightarrow
H}(z,\mu)=\sum_nd_n(z,\mu)\langle0|\mathcal{O}_n^H|0\rangle\; .
\end{eqnarray}
Here, $n$ is the color-spin-orbital quantum number of the heavy
quark pair with null relative momentum.  The short-distance
coefficient $d_n(z,\mu)$, describing the production of heavy quark
pair with appropriate quantum number $n$, is pQCD calculable.
$\mathcal{O}_n^H$ are local four-fermion operators in NRQCD, and
their vacuum expectation values are proportional to the
probabilities of heavy quark pairs with quantum number $n$
hadronizing into quarkonium states. According to NRQCD, in
(\ref{factor-formula}) the short distance sector $d_n(z,\mu)$ can be
computed order-by-order in strong coupling $\alpha_s(2 m_c)$, and
the long distance sector, the matrix elements, can be expanded in
series of the typical relative velocity $v$ of heavy quarks inside
heavy quarkonium \cite{Braatenprd52,nrqcd2}. In principle the
fragmentation function is calculable to any order in $\alpha_s$ and
$v$ as desired; in practice, normally the NLO results are enough for
phenomenological aim. It should be mentioned that in recently there
are discussions about whether the definition of fragmentation
function beyond NLO in strong coupling expansion is complete or not
in the framework of NRQCD \cite{NQS1,NQS2}, whereas it has no
influence on our study in this work.

\begin{figure}[tb]
\begin{center}
\includegraphics[width=14cm]{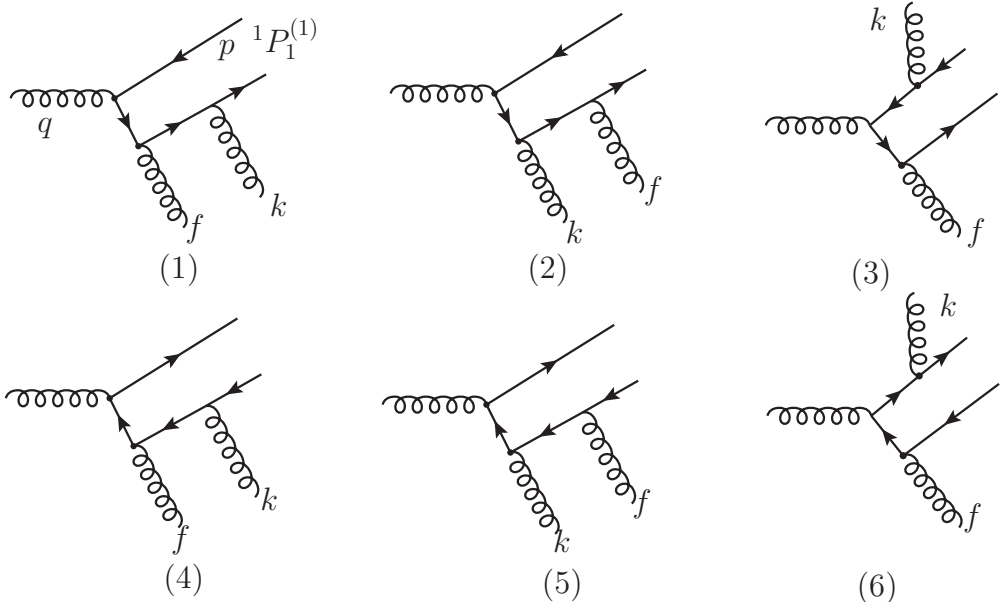}
\caption{Feynmann diagrams of the color-singlet process
$g^*\rightarrow c\bar{c}[^1P_1^{(1)}]+g+g$}\label{fig_singlet}
\end{center}
\end{figure}
\begin{figure}
\begin{center}
\includegraphics[width=9cm]{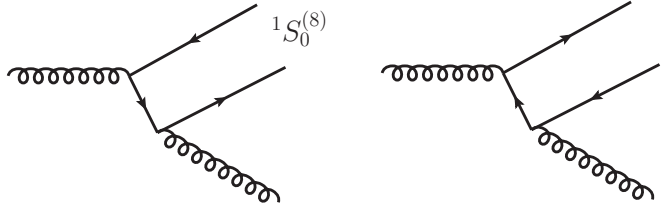}
\caption{Feynmann diagrams of the color-octet process
$g^*\rightarrow c\bar{c}[^1S_0^{(8)}]+g$}\label{fig_octet}
\end{center}
\end{figure}

For process of gluon splitting into $h_c$, there are two classes of
processes at leading order of $\alpha_s$ and $v$. One is the
color-singlet process as shown in Fig.\ref{fig_singlet}, in which
$c\bar{c}$ pair is in spin-singlet, color-singlet and P-wave state
(denoted by $^1P_1^{(1)}$); the other is color-octet process as
shown in Fig.\ref{fig_octet}, in which the quark pair is in
spin-singlet, color-octet and S-wave state($^1S_0^{(8)}$). The full
fragmentation function for $g\rightarrow h_c$ then composes of
color-singlet and color-octet terms, like
\begin{eqnarray}\label{factor-formula-hc}
D_{g\rightarrow
h_c}(z,\mu)=d_1(z,\Lambda)\langle0|\mathcal{O}_1^{h_c}(^1P_1)
|0\rangle+d_8(z)\langle0|\mathcal{O}_8^{h_c}(^1S_0)|0\rangle(\Lambda)\,,
\end{eqnarray}
where the short-distance coefficients $d_1(z,\Lambda)$ and $d_8(z)$
can be obtained through the calculation of Feynmann diagrams
Fig.\ref{fig_singlet} and Fig.\ref{fig_octet}, respectively. The
factorization scale $\Lambda$ is introduced to separate the effect
at short distance of order $1/m_c$ from the one at long distance of
the radius of heavy quarkonium $1/(m_cv)$.
$\langle0|\mathcal{O}_1^{h_c}(^1P_1)|0\rangle$ and
$\langle0|\mathcal{O}_8^{h_c}(^1S_0)|0\rangle$ are matrix elements
of NRQCD operators, scaling as $m_c^5v^5$ and $m_c^3v^5$ according
to velocity-scaling rules in NRQCD. Their dependence on $\Lambda$
can be obtained by renormalization group equations. To leading order
of $\alpha_s(\Lambda)$ they are \cite{nrqcd2},
\begin{eqnarray}\label{rge}
\Lambda\frac{d}{d\Lambda}\langle0|\mathcal{O}_1^{h_c}(^1P_1)
|0\rangle&=&0\,\nonumber\\
\Lambda\frac{d}{d\Lambda}\langle0|\mathcal{O}_8^{h_c}(^1S_0)
|0\rangle&=&\frac{4C_F\alpha_s(\Lambda)}{3N_c\pi
m_c^2}\langle0|\mathcal{O}_1^{h_c}(^1P_1)|0\rangle.
\end{eqnarray}
The last equations can be used to sum up large logarithms of
$m_c/\Lambda$ when the factorization scale are chosen to be much
smaller than $m_c$ \cite{Braatenprd50}. i.e.,
\begin{eqnarray}\label{rgesolution}
\langle0|\mathcal{O}_8^{h_c}(^1S_0)|0\rangle(m_c)=\langle0
|\mathcal{O}_8^{h_c}(^1S_0)|0\rangle(\Lambda)+\frac{4C_F}{3N_c
m_c^2\beta_0}\ln\frac{\alpha_s(\Lambda)}{\alpha_s(m_c)}
\langle0|\mathcal{O}_1^{h_c}(^1P_1)|0\rangle,
\end{eqnarray}
with QCD beta function $\beta_0=(11N_c-2n_f)/$6. In following
analytical results, we will show that $d_8(z)$ is $\Lambda$
independent at leading order of $\alpha_s$. Thus, the dependence on
$\Lambda$ in the factorization formula (\ref{factor-formula-hc})
comes from $\langle0|\mathcal{O}_8^{h_c}(^1S_0)|0\rangle$ and
$d_1(z,\Lambda)$, the same as in the fragmentation functions of
$g\rightarrow \chi_{cJ}$.

Leaving the polarization of fragmenting gluon unsummed, we obtain
the squared amplitudes of color-singlet processes,
\begin{eqnarray}\label{squared-amp}
\sum \mathcal{A}^*_{\mu}\mathcal{A}_{\nu}&=&F_1g_{\mu\nu}+F_2p_\mu
p_\nu +F_3(p_\mu k_\nu
+ k_\mu p_\nu)+F_4k_\mu k_\nu\nonumber\\
&&+(\textsf{terms proportional to}\ q_\mu\ \textsf{or}\ q_\nu)\; .
\end{eqnarray}
Here, $p$, $k$ and $q$ are the momenta of the $c\bar{c}$ pair, final
and initial gluons, and $s=q^2$ is the virtuality of fragmenting
gluon. In the condition $q_0\gg m_c$, which is required to make
fragmentation mechanism a reasonable approximation, the fragmenting
gluon is almost transversely polarized. And under this condition we
can drop terms proportional to $q_\mu$ or $q_\nu$, which are
considered as gauge artifact and suppressed in an appropriate axial
gauge \cite{BraatenLett}. The Lorentz indices in
Eq.(\ref{squared-amp}) imply that there are angle dependence in the
squared amplitudes. After averaging the directions of momentum $p$
and $k$, we get the following replacements,
\begin{eqnarray}
&&p_\mu p_\nu\rightarrow -\frac{1}{2}|\vec{p}_\perp|^2g_{\mu\nu}\,,
\\ &&k_\mu
k_\nu\rightarrow-\frac{1}{2}g_{\mu\nu}\frac{t^2}{y^2-r}
\Big[\,\frac{1}{2}|\vec{p}_\perp|^2(3\cos^2\theta_k-1)+s(y^2-r)(1-\cos^2\theta_k)\Big]\,,\\
&&k_\mu p_\nu+p_\mu k_\nu\rightarrow
-g_{\mu\nu}|\vec{p}_\perp|^2\frac{t\cos\theta_k}{\sqrt{y^2-r}}\; ,
\end{eqnarray}
with
\begin{eqnarray}
\cos\theta_k=\frac{(1+r)/2-t-y+ty}{t\sqrt{y^2-r}}\; , \quad
|\vec{p}_\perp|^2=s(-r+2yz-z^2)\; ,
\end{eqnarray}
where the dimensionless variables are defined,
\begin{eqnarray}
y=\frac{p\cdot q}{s}\; ,\quad t=\frac{k\cdot q}{s}\; ,\quad
r=\frac{4m_c^2}{s}\; ,\quad z=\frac{p^0+p^3}{q^0+q^3}\; .
\end{eqnarray}
Notice that the squared amplitude of Eq.(\ref{squared-amp}) is now
proportional to metric tensor $g_{\mu\nu}$, and the fragmenting
gluon behaves like an on-shell one.

To avoid the presence of infrared divergence while gluons in final
state become soft, we set a lower cutoff $\Lambda$ on the energy of
the final gluons, then the color-singlet coefficient $d_1(z)$
converges and depends on $\Lambda$. After integrating over the
three-body phase space, the color-singlet part of fragmentation
function in Eq.(\ref{factor-formula-hc}) is obtained,
\begin{eqnarray}
d_1(z,\Lambda)&=&\frac{5\alpha_s^3(\mu)}{20736\pi m^5_c}\int^z_0\!\!
dr\int^{(1+r)/2}_{(r+z^2)/2z}dy\frac{1}{(1-y)^4(y^2-r)^2}
\Bigg\{\frac{(1+r^2)-4yz+2z^2}{(1-r)^2}h(y,r)\nonumber\\
&&+\frac{1}{(y-r)^5}\sum_{i=0}^2 z^i
\left((r-y)f_i(y,r)+\frac{g_i(y,r)}{\sqrt{y^2-r}}
\ln\frac{y-r+\sqrt{y^2-r}}{y-r-\sqrt{y^2-r}}\right)\Bigg\}\nonumber\\
&&+\frac{5\alpha_s^3(\mu)}{162\pi
m_c^5}\Big\{-\left[-2z^2+3z+2(1-z)\ln(1-z)\right]
\ln\frac{\Lambda}{m_c}+w(z)\Big\}\,,
\end{eqnarray}
with functions $w(z)$, $h(y,r)$, $f_i(y,r)$ and $g_i(y,r)$ read
{\scriptsize
\begin{eqnarray}
w(z)&=&2(1 - z)\ln^2(1 - z) + [-2 - 3(-2 + z)z + 3(-1 + z)\ln
z]\ln(1 - z) \nonumber\\
&&+
  \frac{1}{6}[2\pi^2(1 - z) + 3z(-7 + 8z) + 3z(-3 + 2z)\ln z]
  - 2(1 - z)Li_2( 1 - z) - (1 - z)Li_2(z)\,,\\
h(y,r)&=&(1+11r-5r^2 + r^3)(1 - 20r + 6r^2 - 4r^3 +
  r^4) + (2 - 20r + 606r^2 - 280r^3 +
   94r^4 - 20r^5 + 2r^6)y \nonumber\\
  && +
 (4 + 468r - 792r^2 + 232r^3 - 44r^4 + 4r^5)
  y^2 + (8 - 1120r + 560r^2 - 96r^3 + 8r^4)
  y^3\nonumber\\
  && + (-240 + 1072r - 208r^2 + 16r^3)y^4 +
 (544 - 448r + 32r^2)y^5 + (-448 + 64r)y^6 +
 128y^7\,,\\
f_0(y,r)&=&r^3 (36 - 15 r + 25 r^2 - 57 r^3 + 59 r^4 - 97 r^5 + 23
r^6 - 7 r^7 + r^8)\nonumber\\
   && + (-164 r^3 + 54 r^4 + 80 r^5 - 34 r^6 + 344 r^7
- 2 r^8 + 12 r^9 -
   2 r^{10}) y\nonumber\\
   &&+ (-82 r^2 + 274 r^3 - 374 r^4 - 46 r^5 - 746 r^6 + 26 r^7 -
   14 r^8 + 2 r^9) y^2\nonumber\\
   && + (-4 r + 488 r^2 + 88 r^3 + 652 r^4 + 824 r^5 -
   564 r^6 - 108 r^7) y^3\nonumber\\
   && + (1 + 33 r - 1277 r^2 - 637 r^3 - 877 r^4 +
   1947 r^5 + 553 r^6 + r^7) y^4 \nonumber\\
   &&+
 (2 - 160 r + 1734 r^2 + 488 r^3 - 2794 r^4 - 1032 r^5 + 66 r^6) y^5\nonumber\\
   && +
 (-28 + 356 r - 1144 r^2 + 2232 r^3 + 644 r^4 - 396 r^5) y^6 +
 (136 - 352 r - 352 r^2 + 400 r^3 + 968 r^4) y^7\nonumber\\
   && +
 (-336 + 80 r - 1136 r^2 - 1264 r^3) y^8 + (448 + 736 r + 992 r^2) y^9 +
 (-384 - 448 r) y^{10} + 128 y^{11}\,,\\
 f_1(y,r)&=&-2yf_2(y,r)\,,\\
 f_2(y,r)&=&
 2\left[r^2 (6 + 8 r - 47 r^2 + 33 r^3 - 24 r^4 + 46 r^5 - 7 r^6 +
r^7)\right. \nonumber\\
   &&+
 (-16 r^2 + 96 r^3 + 56 r^4 + 2 r^5 - 206 r^6 - 10 r^7 - 2 r^8) y\nonumber\\
 && +
 (-16 r - 148 r^2 - 248 r^3 - 74 r^4 + 502 r^5 + 110 r^6 + 2 r^7) y^2 \nonumber\\
   &&+
 (124 r + 470 r^2 + 286 r^3 - 614 r^4 - 234 r^5 + 16 r^6) y^3 \nonumber\\
 &&+
 (-7 - 285 r - 592 r^2 + 320 r^3 + 151 r^4 - 99 r^5) y^4 +
 (34 + 336 r + 204 r^2 + 128 r^3 + 242 r^4) y^5\nonumber\\
 &&\left. +
 (-84 - 220 r - 340 r^2 - 316 r^3) y^6 + (112 + 232 r + 248 r^2) y^7 +
 (-96 - 112 r) y^8 + 32 y^9\right]\,,\\
g_0(y,r)&=&2 \left[r^4 (-9 + r - 8 r^2 - 8 r^3 + r^4 - r^5 - 8 r^6)
+
  (46 r^4 + 10 r^5 + 71 r^6 + 20 r^7 + 4 r^8 + 66 r^9 + 7 r^{10}) y\right.\nonumber\\
  && +
  (28 r^3 - 111 r^4 - 94 r^5 - 148 r^6 - 34 r^7 - 213 r^8 - 20 r^9) y^2\nonumber\\
   && +
  (-148 r^3 + 193 r^4 + 146 r^5 + 150 r^6 + 356 r^7 - 111 r^8 - 26 r^9)
   y^3\nonumber\\
  && + (-28 r^2 + 297 r^3 - 232 r^4 - 120 r^5 - 330 r^6 + 695 r^7 +
    182 r^8) y^4 \nonumber\\
   &&+ (192 r^2 - 166 r^3 + 452 r^4 + 152 r^5 - 1604 r^6 -
    498 r^7 + 16 r^8) y^5 \nonumber\\
  &&+ (-608 r^2 - 590 r^3 - 662 r^4 + 1930 r^5 +
    614 r^6 - 124 r^7) y^6 \nonumber\\
   &&+ (1160 r^2 + 1152 r^3 - 896 r^4 - 112 r^5 +
    408 r^6) y^7\nonumber\\
  && + (8 r - 1152 r^2 - 216 r^3 - 696 r^4 - 744 r^5) y^8 +
  (-104 r + 360 r^2 + 864 r^3 + 816 r^4) y^9\nonumber\\
  &&\left. +
  (8 + 216 r - 376 r^2 - 520 r^3) y^{10} + (-16 - 48 r + 160 r^2)
  y^{11}\right]\,,\\
  g_1(y,r)&=&-2yg_2(y,r)\,,\\
  g_2(y,r)&=&2\left[r^3 (-3 - 5 r + 8 r^2 + 3 r^4 + 13 r^5 + 16 r^6) -
 (-10 r^3 - 14 r^4 + 11 r^5 + 4 r^6 + 84 r^7 + 138 r^8 + 11 r^9) y\right.\nonumber\\
 && -
 (-10 r^2 + 11 r^3 + 114 r^4 + 60 r^5 - 196 r^6 - 495 r^7 - 76 r^8) y^2\nonumber\\
   && -
 (16 r^2 - 183 r^3 - 492 r^4 + 86 r^5 + 940 r^6 + 201 r^7 - 8 r^8) y^3\nonumber\\
 && -
 (16 r + 175 r^2 + 796 r^3 + 640 r^4 - 914 r^5 - 215 r^6 + 62 r^7) y^4\nonumber\\
   && -
 (-96 r - 688 r^2 - 1172 r^3 + 108 r^4 - 76 r^5 - 204 r^6) y^5 \nonumber\\
 &&-
 (218 r + 922 r^2 + 562 r^3 + 502 r^4 + 372 r^5) y^6 +
 (200 r + 472 r^2 + 568 r^3 + 408 r^4) y^7\nonumber\\
 && -
 (-4 + 36 r + 268 r^2 + 260 r^3) y^8 - (8 - 8 r - 80 r^2) y^9\; .
\end{eqnarray}}

Compared to the color-singlet short-distance coefficient, the
calculation for the color-octet one is much more straightforward and
the result does not depend on $\Lambda$,
\begin{eqnarray}
d_8(z)=\frac{5\alpha_s^2(\mu)}{96m_c^3}[-2z^2+3z+2(1-z)\ln(1-z)]\; .
\end{eqnarray}
Now we have the final expression for fragmentation function
$D_{g\rightarrow h_c}(z,\mu)$ at the leading order in strong
coupling constant. If the value for factorization scale is chosen to
be much smaller than $m_c$, we should exploit the renormalization
group equation (\ref{rge}), or its solution (\ref{rgesolution}), to
sum up the logarithm of $\Lambda/m_c$. In this work, the
factorization scale is set to be $m_c$ so that we can avoid the
large logarithms \cite{Braatenprd50}. The value of the scale $\mu$
in $\alpha_s$ are chosen to be $2m_c$, which is the square root of
the minimum value of the virtuality of the fragmenting gluon.

\begin{figure}[htdp]
\begin{center}
\includegraphics[width=14cm]{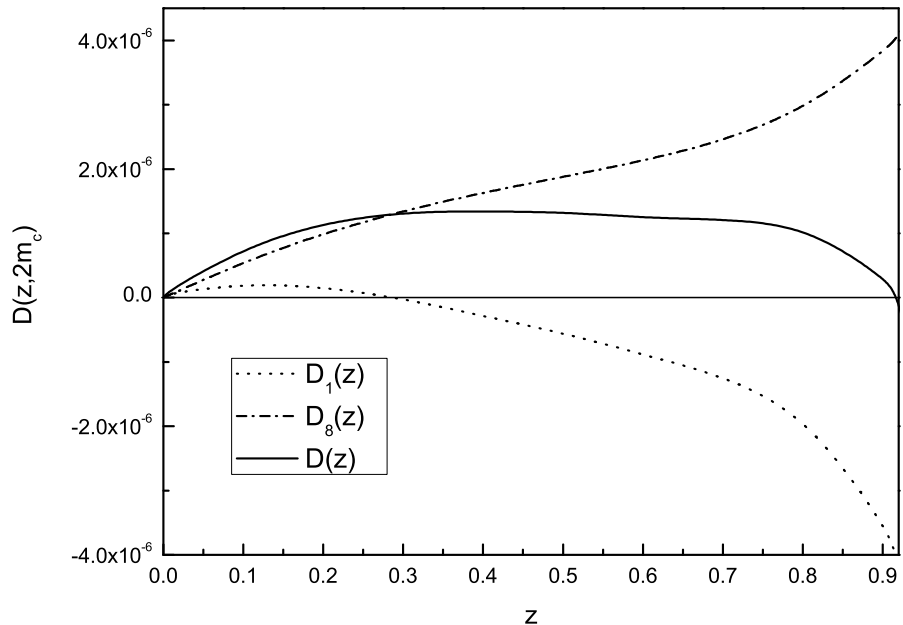}
\caption{The fragmentation function $D_{g\rightarrow h_c}(z,\mu)$ as
a function of $z$ for $\mu=2m_c$ and $\Lambda=m_c$, the dotted line
is the contribution from color-singlet process, the dash-dotted line
is the contribution from color-octet process and the solid line is
the total result. }\label{fig_dz}
\end{center}
\end{figure}

Integrating the fragmentation function $D_{g\rightarrow
h_c}(z,2m_c)$ over the momentum fraction $z$, we get the
fragmentation probability,
\begin{eqnarray}\label{total-jilv}
\int_0^1\!\!\! dz\,D_{g\rightarrow
h_c}(z,2m_c)=\frac{5\alpha_s^2(2m_c)}{96m_c^3}
\Big[\frac{(-54.5)\alpha_s(2m_c)}{216\pi
m_c^2}\langle\mathcal{O}_1^{h_c}\rangle+(0.33)
\langle\mathcal{O}_8^{h_c}\rangle(m_c)\Big]\,.
\end{eqnarray}
To make the total probability positive, the lower bound for the
magnitude of color-octet matrix element at factorization scale
$\Lambda=m_c$ is set to be
\begin{eqnarray}
\langle\mathcal{O}_8^{h_c}\rangle(m_c) > \frac{3\alpha_s(2m_c)}{4\pi
m_c^2}\langle\mathcal{O}_1^{h_c}\rangle \; .
\end{eqnarray}
For numerical evaluation, the input parameters are taken as follows:
$m_c=m_{h_c}/2=1.78$GeV, $\alpha_s(2m_c)=0.26$,
$\langle0|\mathcal{O}_1^{h_c}(^1P_1)|0\rangle=0.32\text{GeV}^5$
\cite{O1value}, which enables
$\langle0|\mathcal{O}_8^{h_c}(^1S_0)|0\rangle(m_c) > 6.3\times
10^{-3}\text{GeV}^3$,
$\langle0|\mathcal{O}_8^{h_c}(^1S_0)|0\rangle(m_c)=9.8\times
10^{-3}\text{GeV}^3$ \cite{O8value1,O8value2}. With above inputs the
fragmentation probability in Eq.(\ref{total-jilv}) is found to be
about $7.1\times 10^{-7}$. Compared with the probabilities of gluon
fragmenting into $\chi_{c0}$, $\chi_{c1}$, and $\chi_{c2}$, which
are $0.4\times 10^{-4}$, $1.8\times 10^{-4}$ and $2.4\times 10^{-4}$
\cite{Braatenprd50}, the value of gluon fragmentation to $h_c$ is
smaller by two to three orders, and it is even less than the
color-singlet probability of process $g^*\rightarrow J/\psi gg$ by
an order \cite{Braatenprd52}. The $z$ dependence of the
fragmentation function at $\mu=2m_c$ and $\Lambda=m_c$ is shown in
Figure \ref{fig_dz}.

In conclusion, we have computed in this work the fragmentation
function of gluon to P-wave spin singlet quarkonium $h_c$. The
analytic expression is infrared safe while both color-singlet and
-octet processes are taken into account. It is found that the
fragmentation probability of a high energy virtual gluon splitting
into $h_c$ is about $7.1\times 10^{-7}$. The obtained analytic
fragmentation function can be employed to the study of $h_c$
hadroproduction, especially in the large transverse momentum region.
Finally, it is worthy to note that result in this work can be
readily applied to the study of $h_b$ physics.

%%%%%%%%%%%%%%%%%%%%%%%%%%%%%%%%%%%%%%%%%%%%%%%%%%%%%%%%%%%%%%%%%%%%%
\vspace{.9cm} {\bf Acknowledgments} \vspace{.3cm}

This work was supported in part by the National Natural Science
Foundation of China(NSFC) under the grants 10935012, 10928510,
10821063, 10775179 and by CAS Key Projects (KJCX2-yw-N29,
H92A0200S2).

%%%%%%%%%%%%%%%%%%%%%%%%%%%%%%%%%%%%%%%%%%%%%%%%%%%%%%%%%%%%%%%%%%%%%

\end{document}